\documentclass[10pt,conference]{IEEEtran}
\IEEEoverridecommandlockouts

\usepackage{cite}
\usepackage{amsmath,amssymb,amsfonts}
\usepackage{physics}
\usepackage{bbm}
\usepackage{algorithmic}
\usepackage{graphicx}
\usepackage{textcomp}
\usepackage{xcolor}
\usepackage{tabu}
\usepackage{mathtools}
\usepackage{listings}
\usepackage{url}

\definecolor{erblue}{HTML}{0082F0}
\definecolor{erorange}{HTML}{FF8C0A}
\definecolor{ergreen}{HTML}{0FC373}
\definecolor{erpurple}{HTML}{AF78D2}
\definecolor{eryellow}{HTML}{FAD22D}
\definecolor{erred}{HTML}{FF3232}

\newcommand{\MMD}{\operatorname{MMD}}

\def\BibTeX{{\rm B\kern-.05em{\sc i\kern-.025em b}\kern-.08em
    T\kern-.1667em\lower.7ex\hbox{E}\kern-.125emX}}
\begin{document}

\title{On the learning abilities of photonic continuous-variable Born machines}

\author{
    \IEEEauthorblockN{
        Zoltán Kolarovszki\IEEEauthorrefmark{1}\IEEEauthorrefmark{2},
        Dániel T. R. Nagy\IEEEauthorrefmark{1}\IEEEauthorrefmark{2},
        Zoltán Zimborás\IEEEauthorrefmark{1}\IEEEauthorrefmark{2}\IEEEauthorrefmark{3}
    }
    \IEEEauthorblockA{\IEEEauthorrefmark{1}\textit{Wigner Research Centre for Physics}, Budapest, Hungary}
    \IEEEauthorblockA{\IEEEauthorrefmark{2}\textit{Eötvös Loránd University}, Budapest, Hungary}
    \IEEEauthorblockA{\IEEEauthorrefmark{3}\textit{Algorithmiq Ltd}, Helsinki, Finland\\
    Email: \{kolarovszki.zoltan, nagy.dani, zimboras.zoltan\}@wigner.hun-ren.hu
    }
}

\maketitle

\begin{abstract}
    This paper investigates photonic continuous-variable Born machines (CVBMs), which utilize photonic quantum states as resources for continuous probability distributions.
    Implementing exact gradient descent in the CVBM training process is often infeasible, bringing forward the need to approximate the gradients using an estimator obtained from a smaller number of samples, obtaining a quantum stochastic gradient descent (SGD) method.
    In this work, the ability to train CVBMs is analyzed using stochastic gradients obtained using relatively few samples from the probability distribution corresponding to homodyne measurement.
    The main obstacle to this analysis is that classically simulating CVBMs and obtaining samples is a demanding task, while a large number of iterations are needed to achieve convergence.
    The present research is enabled by a novel strategy to simulate homodyne detections of generic multimode photonic states using a classical computer.
    With this approach, a more comprehensive study of CVBMs is made possible, and the training of multimode CVBMs is demonstrated with parametric quantum circuits considerably larger than in previous articles.
    More specifically, we use the proposed algorithm to demonstrate learning of multimode quantum distributions using CVBMs.
    Moreover, successful CVBM trainings were demonstrated with the use of stochastic gradients.
\end{abstract}


\section{Introduction}

    \IEEEPARstart{I}n the last decade, significant advancements have been made in the development of quantum computer prototypes. Notably, there has been a growing emphasis on photonic quantum computers, driven by recent demonstrations of potential photonic quantum advantage schemes~\cite{photonicadvantage1, zhong2021phase,madsen2022quantum} and the refinement of viable fault-tolerant quantum computation techniques~\cite{bartolucci2023fusion,bombin2023logical,bourassa2021blueprint}. Among the feasible near-term applications of photonic quantum computers, those that use parametrized quantum circuits (PQCs) have gained much attraction recently, since they can be regarded as the quantum counterparts of classical neural networks~\cite{Killoran_2019,Miatto_2020}.
    In particular, due to the continuous nature of the optical phase space, generative modeling has become a promising use case for photonic PQCs.
     
    The \textit{continuous-variable Born machine} (CVBM), introduced in Ref.~\cite{cepaite2022}, is an instance of generative quantum machine learning (QML) models.
    The CVBM algorithms employ photonic quantum states for generating probability distributions through their homodyne measurement, where the probability distributions can be computed using Born's rule~\cite{Coyle_2020}. In this setup, the quantum states may be adjusted using a PQC, thus modifying the output distribution. Therefore, one needs to use classical optimization techniques to tune the parameters of the PQC to approximate some target distribution.
    The optimization of the parameters of the Born machines can be done using the \textit{parameter-shift rules} for certain quantum gates; these rules estimate the gradient from samples of states arising from PQCs with different shifted parameters~\cite{Mitarai_2018,Schuld_2019}.

    Let us note, however, that exact gradient descent is infeasible due to the huge number of measurements required for sufficiently precise evaluation. Instead, recent observations suggest that using a finite number of measurements effectively implements stochastic gradient descent (SGD), replacing exact partial derivatives with estimators obtained from a relatively small number of samples~\cite{Sweke_2020,Harrow_2021}.

    In this paper, we will investigate the learning abilities of CVBMs, especially studying the experimentally feasible case when only a smaller number of samples are used to estimate the gradient, and thus the target distribution is reached through a quantum SGD method.

    In general, both for classical and quantum neural network training, many iterations are needed to achieve convergence. Since simulating a CVBM circuit and obtaining samples is a notoriously demanding task for a classical computer, previous numerical studies exclusively resorted to single-mode systems~\cite{cepaite2022}.

    The present investigation is enabled by a novel and efficient approach developed by us to simulate classically homodyne detections of generic multimode photonic states. Our approach to sampling from multiple modes is made possible because of two reasons:
    (1) By taking samples with an already-known weak simulation strategy called \textit{mode-by-mode sampling};
    (2) by using \textit{inverse transform sampling} instead of the previous discrete approximation of the single-mode probability distributions.

    The structure of the paper is as follows: in Sec.~\ref{sec:basics}, we lay out the basics of homodyne measurements and CVBMs. Sec.~\ref{sec:algorithm} provides an algorithm for homodyne sampling of generic states represented in the truncated Fock space. Finally, in Sec.~\ref{sec:results} we show some examples of CVBM training with a higher number of shots, weights, and modes than in previous investigations.

\section{Basics} \label{sec:basics}
    \subsection{Homodyne measurement}
        In this section, we provide a brief description of homodyne measurements in photonic quantum computing. For more details on homodyne detection, see Ref.~\cite{serafini}.

        Homodyne detection corresponds to the measurement of the quadratures in the optical phase space. The homodyne detection POVM can be formulated by considering the quadrature operators
        $
            \hat{x}_\phi = \cos \phi \, \hat{x} + \sin \phi \, \hat{p}
        $,
        where the $\hat{x}$ and $\hat{p}$ operators are the position and momentum operators, respectively (also called quadrature operators). By adjusting the angle $\phi$, one can sample from the phase space along different directions. $\phi \equiv 0$, corresponds to the position distribution, while $\phi \equiv \pi / 2$ corresponds to the momentum distribution. In this paper, we set $\phi \equiv 0$ consistently, without loss of generality.
        The projectors in the homodyne detection POVM can be written using the (generalized) eigenvectors of $\hat{x}_\phi$ denoted as $\ket{\hat{x}_\phi}$. Using Born's rule, one can obtain the probability of the detection at position $x$ at angle $\phi$ using the projectors
        $
            \ket{\hat{x}_\phi}\!\bra{\hat{x}_\phi}.
        $

        For practical purposes, it is convenient to express photonic states as a density matrix in the Fock basis for simulations, where some Fock space truncation is imposed. To calculate the position distribution from the density matrix, we can use the well-known formula for the quantum harmonic oscillator wave functions ($\hbar = 1$):
        \begin{equation} \label{eq:wave_function}
            \bra{m_i} \ket{x_i} =: \psi_i(x_i) = \frac{ e^{-\frac{x^2}{2}} }{\sqrt{2^n n! \sqrt{\pi}}} H_n(x),
        \end{equation}
        where $H_n$ are the Hermite polynomials defined by
        $
            H_n(x) = (-1)^n e^{x^2}\frac{\dd^n}{\dd x^n}e^{-x^2}
        $.

    \subsection{Training CVBMs}
        In this section, we give a brief overview of the training of CVBMs. For a more detailed description, see Ref.~\cite{cepaite2022}.

        A CVBM generates statistics from the probability distribution specified by the homodyne measurement, i.e., from the position distribution of the state, which is parametrized by a PQC. This probability distribution can be written as
        \begin{equation}
            p(\vec{x}) = |\! \bra{\vec{x}} U(\vec{w}) \ket{\psi_0} \! |^2, 
        \end{equation}
        where $U(w)$ corresponds to a continuous-variable quantum neural network, $\vec{w}$ represents the neural network weights, and $\ket{\psi_0}$ is the input state. In machine learning, the parametrized quantum circuit $U(\vec{w})$ serves as the generative model, and the state $\ket{\psi(w)} \coloneqq U(\vec{w}) \ket{\psi_0}$, when sampled with a fixed measurement, generates the probability distribution of interest.

        In our work, we follow Ref.~\cite{cepaite2022} and use the maximum mean discrepancy (MMD) to calculate the loss for the training. Given two probability distributions $P$ and $Q$, the MMD of these distributions can be given by
        \begin{equation}
            \operatorname{MMD}(P,Q) = \underset{\substack{x \sim P\\y \sim P}}{\mathbb{E}}[k(x, y)]
            +
            \underset{\substack{x \sim Q\\y \sim Q}}{\mathbb{E}}[k(x, y)]
            -
            2 
            \underset{\substack{x \sim P\\y \sim Q}}{\mathbb{E}}[k(x, y)],
        \end{equation}
        where $k$ is a kernel function. Throughout this paper, we used the classical Gaussian kernel $k$ defined by
        $
            k(x , y ) = e^{-||x -y ||^2  / 2\sigma^2}.
        $
        Moreover, we consistently set $\sigma \equiv 1$.
        However, MMD could only be estimated using the samples provided by the photonic circuit, and hence we can only calculate the (unbiased) MMD estimator given by
        \begin{align}
        \begin{split}
        \label{eq:mmd_estimator}
            \operatorname{MMD}(P,Q)
            &\approx
            \frac{1}{M(M-1)}\sum_{i\neq j}^M k(x_i, x _j)
            \\&+ \frac{1}{N (N-1)}\sum_{i\neq j}^N k( y_i, y_j)
            - \frac{2}{MN}\sum_{i,j}^{M, N} k(x_i,y_j),
        \end{split}
        \end{align}
        where $\vec{x}$ and $\vec{y}$ are samples from $P$ and $Q$, respectively.
    
        Given a distribution $P$ and target distribution $Q$, the loss function can be given by the MMD as
        $
            L_{\mathrm{MMD}}(P) \coloneqq \mathrm{MMD}(P, Q).
        $
        For training the CVBM, however, one also needs to approximate the gradients of the loss function with respect to the circuit parameters.
        This can be achieved by using the parameter-shift rule for photonic gates. The used photonic gates are tabulated in Table~\ref{table:gate_definitions} for completeness, and the analytic expressions for the parameter-shift rules for Gaussian gates are provided in Ref.~\cite{cepaite2022}, which are also collected in Table~\ref{table:parameter_shift_rules} for completeness.
        Considering a gate $G$ with corresponding weight $w$, the estimator of the loss gradient can be written as
        \begin{align}
        \begin{split} \label{eq:mmd_grad}
            \partial_{w} L_{\mathrm{MMD}}(\vec{w}) &\approx 
                \frac{m_G}{RM} \sum_{i,j}^{R, M} k(a_i, x_j)
                - \frac{m_G}{SM} \sum_{i,j}^{S, M} k(b_i,x_j)
                \\&
                - \frac{m_G}{RN} \sum_{i,j}^{R, N} k(a_i, y_j)
                + 
                \frac{m_G}{SN} \sum_{i,j}^{S, N} k(b_i, y_j),
        \end{split}
        \end{align}
        where $m_G$ is the multiplier corresponding to the gate $G$ according to Table~\ref{table:parameter_shift_rules} and $\vec{a}$ and $\vec{b}$ are sampled from the positively and negatively shifted circuits with shifts $\pm s_G$, respectively~\cite{Liu_2018}. Throughout the paper, $s_{S} = 1.0$ and $s_{D} = 1.0$ are used in the parameter-shift rule of the squeezing and displacement gates, respectively.

        \begin{table}[ht]
            \centering
            \caption{Definitions photonic quantum gates.}
            \renewcommand{\arraystretch}{2.0}
            \begin{tabular}{c c c}
            \textbf{Gate} & \textbf{Unitary operator} \\
                \hline \\ [-6.0ex]
                Phaseshift gate & $R_j(\phi) \coloneqq \exp(i \phi a_j^\dagger a_j)$ \\
                Beamsplitter gate & $BS_{jk}(\theta, \phi) \coloneqq \exp(\theta ( e^{i \phi} a_j a_k^\dagger - e^{-i \phi} a_j^\dagger a_k))$ \\
                Displacement gate & $D_j(\alpha) \coloneqq \exp(\alpha a_j^\dagger - \alpha^* a_j)$ \\
                Squeezing gate & $S_j(z) \coloneqq \exp(\bar{z} a_j^2 - z a_j^{\dagger 2} )$ \\
                Cubic phase gate & $V(\gamma) \coloneqq \exp(i \hat{x}^3 \frac{\gamma}{3 \hbar})$ \\
                Cross-Kerr gate & $CK_{jk}(\kappa) \coloneqq \exp \left (
                    i \xi n_j n_k
                \right )$ \\
                \hline
            \end{tabular}
            \label{table:gate_definitions}
        \end{table}
    
        \begin{table}[ht]
            \centering
            \caption{Parameter-shift rules for Gaussian gates.}
            \renewcommand{\arraystretch}{1.4}
            \begin{tabular}{c c c}
            \textbf{Gate $G$} & \textbf{Parameter-shift $s_G$} & \textbf{Multiplier $m_G$}\\
                \hline \\ [-3.5ex]
                $R(\phi)$ & $\pi / 2$ & $1$ \\
                $BS(\theta, \phi)$ & $\pi / 2$ & $1$ \\
                $D(\alpha)$ & $s_D \in \mathbb{R}$  & $s_D^{-1}$ \\
                $S(r)$ & $s_S \in \mathbb{R}$ &  $\sinh (s_S)^{-1}$ \\
                \hline
            \end{tabular}
            \label{table:parameter_shift_rules}
        \end{table}

\section{Classical simulation algorithm} \label{sec:algorithm}
    In this section, we provide a weak simulation strategy to perform homodyne sampling from a generic multimode photonic state with a given density matrix, i.e., we sample from the position distribution while avoiding explicitly calculating the distribution. To our knowledge, this approach is novel and has not been investigated yet. The strategy described in this section is published in the Piquasso photonic quantum computer simulator as \lstinline|pq.HomodyneMeasurement|~\cite{kolarovszki2024piquasso}.

    \begin{figure}[ht]
        \centering
        \includegraphics[width=0.45\textwidth]{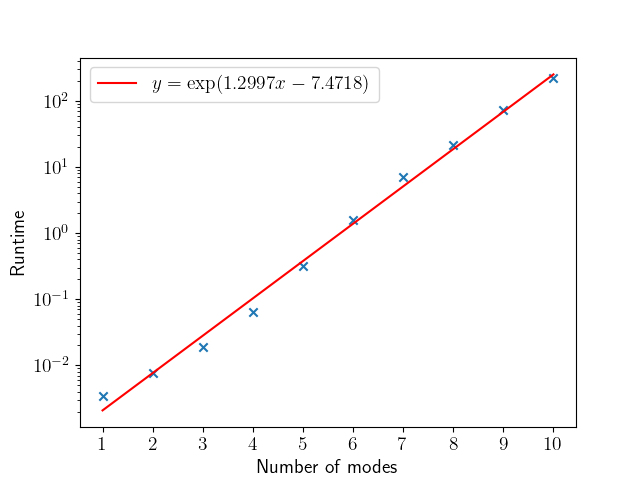}
        \caption{The benchmarks were run using a fixed Fock space cutoff $7$, and $100$ samples were taken for each iteration. Up to $6$ number of modes, the results were averaged from $100$ iterations with $10$ warmup iterations, and for a higher number of modes, only a single iteration was executed with a single warmup run. The benchmark was executed on an Intel(R) Core(TM) i7-10610U CPU @ 1.80GHz architecture.}
        \label{fig:benchmark}
    \end{figure}

    A density matrix on the $d$-mode Fock space $\rho$ can be expanded in the occupation number basis as
    \begin{equation}
        \rho = \sum_{\vec{n} \geq 0, \vec{m} \geq 0} \rho_{\vec{n}, \vec{m}} \ket{\vec{n}}\!\bra{\vec{m}},
    \end{equation}
    For practical purposes, the density matrix is truncated so that the sum of the occupation numbers should not reach a prescribed threshold number $c$. This method is thoroughly explained by the authors in Ref.~\cite{kolarovszki2024piquasso}. Consequently, all summations over the occupation numbers $\vec{n}$ should be understood with the condition $\sum_{i=1}^d n_i < c$ in this paper.
    Furthermore, suppose that the density matrix $\rho$ is normalized after the truncation.

    We can calculate the position distribution in the phase space using Born's rule and the POVM for the homodyne measurement as
    \begin{equation} \label{eq:jointprob}
        p(\vec{x})
        =
        \Tr [ \rho  \ket{\vec{x}}\!\bra{\vec{x}} ]
        = 
        \sum_{\vec{n} \geq 0, \vec{m} \geq 0} \rho_{\vec{n}, \vec{m}} \bra{\vec{x}} \ket{\vec{n}}\!\bra{\vec{m}} \ket{\vec{x}},
    \end{equation}
    where
    $
        \bra{\vec{m}} \ket{\vec{x}} = \prod_{i=1}^d  \bra{m_i} \ket{x_i}.
    $
    To sample from this distribution approximately, one can discretize the probability density and take samples from the discretized distribution. However, the precision of this approach heavily depends on the resolution. Moreover, as the dimension gets larger, this approach becomes intractable, because the number of bins required to approximate the probability distribution within a given error bound grows exponentially with the dimension.

    To sample from the probability distribution $p(\vec{x})$ in multiple modes, we followed the mode-by-mode sampling approach, already used in the simulation of Gaussian Boson Sampling (see, e.g., Ref.~\cite{quesada2020exact}). More concretely, the strategy is to acquire a sample on each mode iteratively by calculating the conditional probabilities.
    For the $i$-th mode, one can sample $s_i$ using the previous samples $s_1, \dots, s_{i-1}$ by using the following density matrix:
    \begin{equation}
        \rho^{(i)} = \frac{
            \Tr_{i,\dots,d}\left[
                \left(
                    \bigotimes_{j=1}^{i-1} \ket{s_i}\!\bra{s_i} \otimes \mathbbm{1}
                \right) \Tr_{i+1,\dots,d} \rho
            \right]
        }{
            \Tr_{i,\dots,d}\left[
                \left(
                    \bigotimes_{j=1}^{i-1} \ket{s_i}\!\bra{s_i}
                \right) \Tr_{i+1,\dots,d} \rho
            \right]
        }.
    \end{equation}
    By direct calculation, it can be shown that this density matrix yields the probability conditioned on the previous samples when used with Born's rule, i.e.,
    \begin{align} \begin{split}
        p_i(x) &\coloneqq \frac{p_{1\dots i}(x_1=s_1, \dots, x_{i-1} = s_{i-1}, x)}{p_{1\dots (i-1)}(x_1=s_1, \dots, x_{i-1} = s_{i-1})}
        \\&= \Tr [ \ket{x}\!\bra{x} \rho^{(i)} ].
    \end{split}
    \end{align}
    Using Eq.~\eqref{eq:wave_function}, we can decompose $p_i(x)$ as
    \begin{equation} \label{eq:p_1}
        p_i(x) = Q_i(x) e^{-x^2},
    \end{equation}
    where $Q_i(x)$ is a polynomial assuming that the Fock space cutoff $c$ is imposed. This polynomial can explicitly be written as
    \begin{equation}
        Q_i(x) = \sum_{n, m=0}^{c-1} \rho_{n,m}^{(i)} \frac{  H_n(x) H_m(x)  }{\sqrt{2^{n+m} n! m! \pi}},
    \end{equation}
    according to the overlap formula Eq.~\eqref{eq:wave_function}.
    
    To sample from this distribution, one can use the \textit{inverse transform sampling} method~\cite{vogel02}, i.e., knowing the cumulative distribution function $F_i(t) \coloneqq \int_{-\infty}^t p_i(x) \dd x$, we can generate a random number $\alpha_i$ from the uniform distribution on $[0, 1]$, and search for the solution of the equation 
    \begin{equation} \label{eq:root_finding}
        F_i(t) = \alpha_i.
    \end{equation}
    The desired sample $s_i$ in the $i$st mode is the solution of this equation. However, the primitive function of $p_i$ is not analytic, and hence Eq.~\eqref{eq:root_finding} is difficult to solve explicitly in general.
    This implies the need to use a root-finding algorithm to establish a solution numerically. It should be emphasized, that one needs to use a root-finding algorithm with guaranteed convergence. In this work, we used Brent's 
    method~\cite{brent1971} implemented in SciPy~\cite{scipy}, because it has a good convergence rate and is guaranteed to converge. To efficiently perform a step during root finding, one has to make sure that the function evaluations are sufficiently fast, hence it is required to put $F_1$ in a form that is efficient to evaluate. Integrating Eq.~\eqref{eq:p_1} by parts and using the formula
    \begin{equation}
        \frac{\dd^k}{\dd x^k} H_n(x) = 2^k k! \binom{n}{k} H_{n-k}(x),
    \end{equation}
    it can be shown that the cumulative distribution function $F_i$ is
    \begin{equation}
        F_i(t) = \frac{\erf(t)+1}{2}
        -
        e^{-t^2} A_{\rho^{(i)}}(t),
    \end{equation}
    where $\erf$ is the error function and $A_{\rho^{(i)}}$ is a polynomial of the form
    \begin{align}
        A_{\rho^{(i)}}(t) \coloneqq \sum_{n,m=0}^c
         \rho_{n,m}^{(i)}
            B_{n,m}(t),
    \end{align}
    where
    \begin{align} \begin{split}
        B_{n,m}
        &\coloneqq
        \frac{n!\,
            C_{n,m}
            +
            (1-\delta_{n,m}) n!\, 2^n H_{m-n-1}
        }{
            \sqrt{2^{n+m} n!m! \, \pi}
        },
        \\
        C_{n,m} &\coloneqq \sum_{k=0}^{n-1}  \frac{2^k}{(n-k)!} H_{n-k} * H_{m-k-1},
    \end{split}
    \end{align}
    where $*$ denotes the discrete convolution.
    Hence, sampling from $F_i$ is relatively efficient, since the calculation of the polynomial $A_{\rho^{(i)}}(t)$ can be performed in advance of the root-finding algorithm. The error function can be evaluated approximately with an efficiently computable rational approximation provided by Abramowitz and Stegun in Ref.~\cite{AbraSteg72}. In Piquasso, the approximation (7.1.25) is implemented.

    Strictly speaking, the coefficient of the polynomials $B_{n, m}$ and $H_n$ are independent of the density matrix $\rho$ and hence can be calculated in advance. Moreover, when sampling on a mode $i \in [d]$, it is also beneficial to calculate the coefficients of the polynomial $A_{\rho^{(i)}}$ before starting the root-finding algorithm. This way, each step in the root-finding algorithm is just a simple function evaluation. Importantly, we know that $\deg A_{\rho^{(i)}} = 2c-1$, hence the complexity of the evaluation only scales linearly in the Fock space cutoff. However, the number of evaluations in the root-finding algorithm heavily depends on the distribution.

    Note, that the denominator in $\rho^{(i)}$ is not needed to be calculated explicitly, since $\Tr \rho^{(i)} = 1$.
    However, calculating the numerator of $\rho^{(i)}$ gets intractable as the number of modes $d$ gets larger, since $\rho^{(i)}$ needs to be contracted from $\rho^{(1\dots i)}$, which has dimension $n_i \coloneqq \binom{i+c-1}{i}$  . The contraction requires a loop with $\mathcal{O}(n_i^2)$ steps. Hence, taking a single sample using this algorithm has a (classical) computational complexity upper bound of
    $        \Tilde{\mathcal{O}}\left(
            e^{
                2 (d + c - 1) \log(d+c-1)
            }
        \right),
    $
    where we have bounded the sampling complexity of the $i$-th modes by the complexity of the $d$-th mode. Due to this upper bound, the observed exponent might have a lower actual slope than $2$.
    To better approximate the exponent, the algorithm has been executed for an increasing number of modes with a fixed cutoff, and the results were visualized in Fig.~\ref{fig:benchmark} on a logarithmic scale. In this benchmark, the natural logarithm of the runtimes was fitted with a linear equation, assuming that the logarithm in the exponent is negligible. From this regression, it could be concluded that the observed exponent is around $1.2997$, much lower than the upper bound of $2$.

    \begin{figure}[h]
        \centering
        \includegraphics[width=0.45\textwidth]{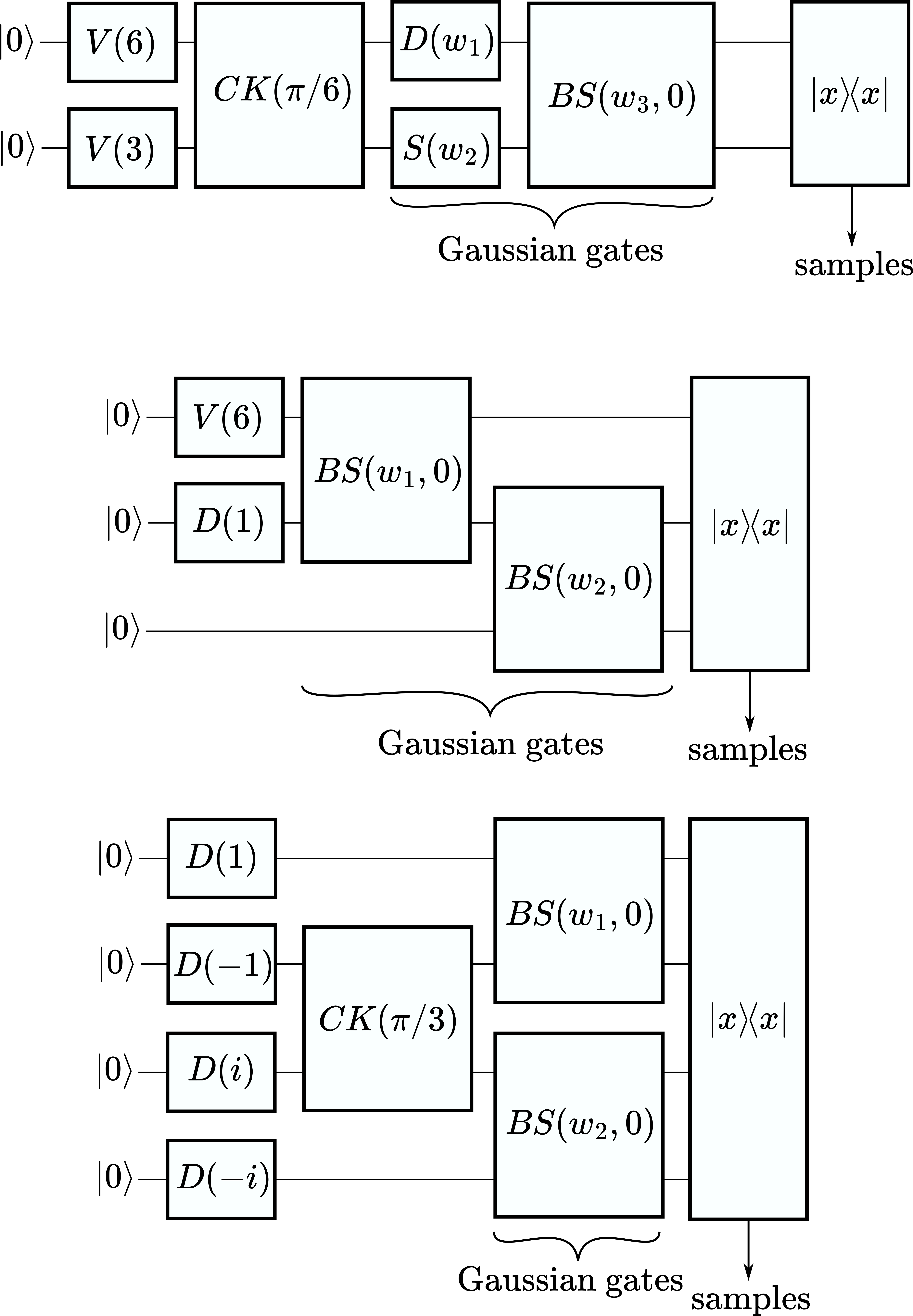}
        \caption{Setup for the $2$-, $3$- and $4$-mode Born machine examples, where the circuit is differentiated with respect to the $w_1, w_2, w_3$ weights.}
        \label{fig:circuits}
    \end{figure}

\vspace*{8mm}
\section{Results} \label{sec:results}
    
    \begin{figure*}[t]
        \centering
        \includegraphics[width=\textwidth]{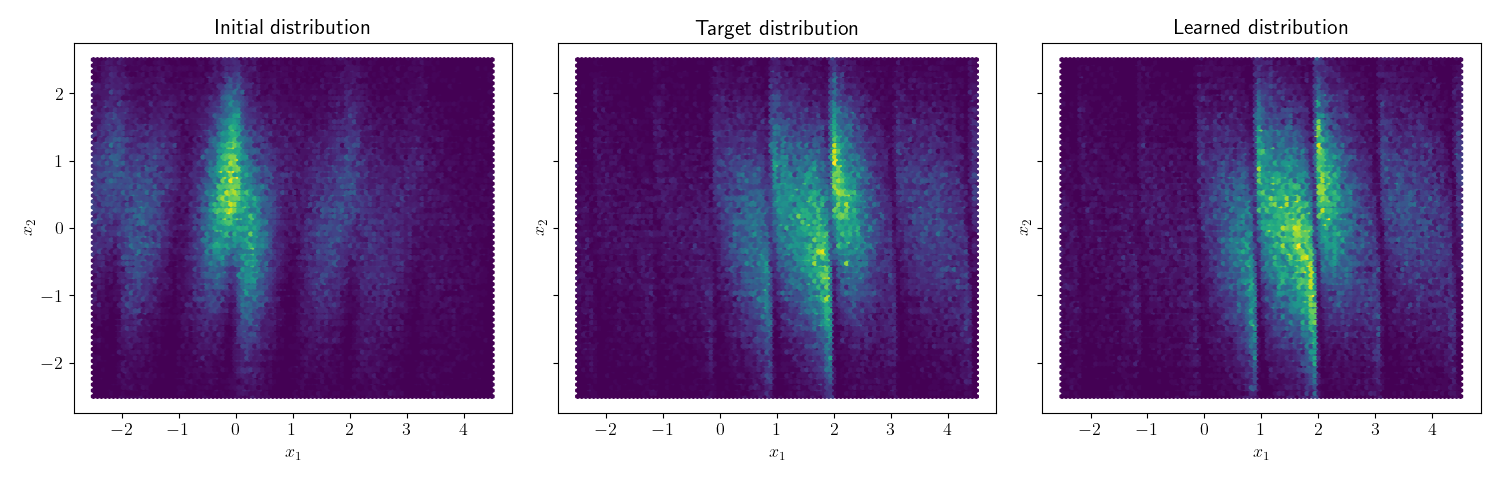}
        \caption{The initial, target, and learned distribution corresponding to the setup in Fig.~\ref{fig:circuits}, sampled from the joint probability distribution from Eq.~\eqref{eq:jointprob}, using 100.000 samples, and visualized via a heatmap. The learned distribution is calculated from the weights corresponding to the minimum loss during the training procedure.}
        \label{fig:distributions_two_mode}
    \end{figure*}

    In this section, we investigate CVBMs using the classical weak simulation algorithm of the homodyne measurement described in Sec.~\ref{sec:algorithm}. This calculation enables us to increase the system size and number of shots and weights. In our study, we focus on quantum distributions, which are probability distributions corresponding to measurements of quantum states. Moreover, we train CVBMs to learn probability distributions of non-Gaussian states, since homodyne detection of Gaussian states is feasible to simulate on a classical computer~\cite{Papoulis}.

    We use the parameter-shift rules described in Sec.~\ref{sec:basics} during the training process. For non-Gaussian gates, no analytic formula is known for the parameter-shift rules, hence, one has to resort to an approximation of the gradient using small shifts in the gate parameters. This method requires a vast amount of samples to correctly approximate the gradient, hence, we only consider cases where the weight corresponds to Gaussian gates. Moreover, it should be noted, that non-Gaussian gates make the parameter-shift rule inapplicable to Gaussian gates appearing before the said non-Gaussian gate. For these reasons, we only vary circuit parameters after all non-Gaussian gates.
    
    For the training, we used the Adam optimizer, which is a stochastic gradient-based optimizer~\cite{kingma2017adam}, with decay rates $\beta_1 = 0.9$ and $\beta_2 = 0.999$ and learning rate $0.01$. The weights are always initialized from $0.0$, and the seed sequence $12345$ has been used for obtaining the samples from Piquasso. Moreover, all the simulations were executed using $\hbar = 2.0$.\footnote{The source code is published at \url{https://github.com/Budapest-Quantum-Computing-Group/piquasso_cvbm}}

\begin{figure}[t]
        \centering
        \includegraphics[width=0.45\textwidth]{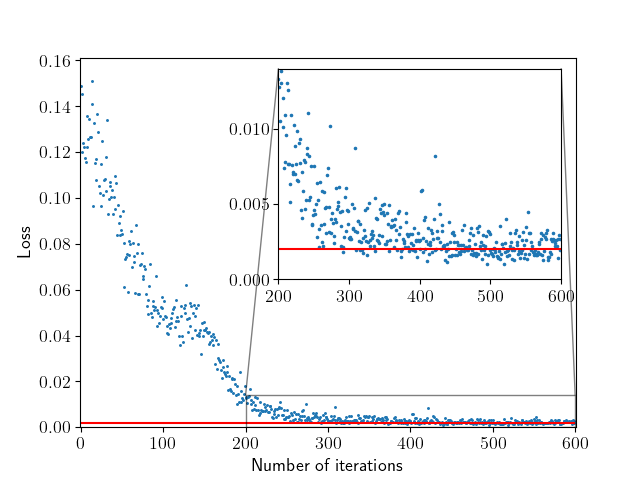}
        \caption{The training loss curve for the $2$-mode CVBM specified in Fig.~\ref{fig:circuits}. The blue dots indicate the losses at each iteration. The red line indicates a mean value of $L_{\MMD}(Q)$ averaged from $100$ values, where $Q$ is the target distribution.}
        \label{fig:losses_2_mode}
    \end{figure}

\begin{figure}[t]
        \centering
        \includegraphics[width=0.45\textwidth]{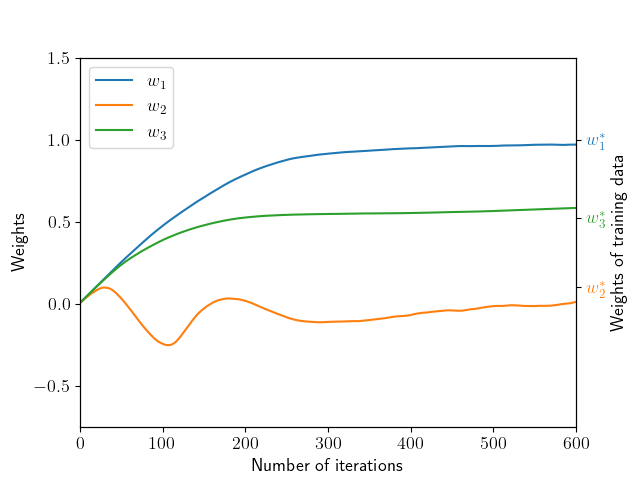}
        \caption{The values of the weights $w_1, w_2, w_3$ during the learning process.
        For the learning, $1000$ shots were used to approximate the MMD loss and its gradient. The learning rate was $0.005$ and the Fock space cutoff was $10$.
        The $600$ iterations were executed in $1739$ s on an Intel(R) Core(TM) i7-10610U CPU @ 1.80GHz architecture.}
        \label{fig:weights_2_mode}
    \end{figure}

\begin{figure}[h]
        \centering
        \includegraphics[width=0.45\textwidth]{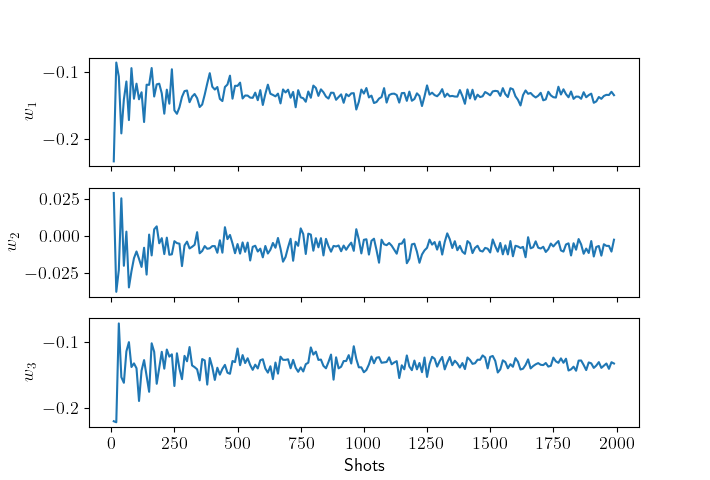}
        \caption{The estimations of the initial gradients in the two-mode CVBM setup described in Fig.~\ref{fig:circuits} in terms of the number of shots taken.}
        \label{fig:gradient_estimates}
    \end{figure}

    We successfully trained a $2$-mode circuit consisting of cubic phase gates and a cross-Kerr gate, with $3$ Gaussian gates followed, as seen in Fig.~\ref{fig:circuits}. The target distribution was generated with weights
    $
        w_1^* = 1.0, w_2^* = 0.1$ and $w_3^* = \pi / 6.
    $
    The learned and the target distributions are compared in Fig.~\ref{fig:distributions_two_mode}, which shows that the learned and target distributions match visually.
    The losses during the training process are shown in Fig.~\ref{fig:losses_2_mode}, and the values of the weights during the training process are visualized in Fig.~\ref{fig:weights_2_mode}. These figures also demonstrate that the training procedure was indeed successful. Moreover, we also investigated the effects of the number of shots on the gradient, which is visualized in Fig.~\ref{fig:gradient_estimates}. While the CVBM training converges well, the graph shows that using $1000$ shots does not approximate the MMD gradients well. This could indicate, that it might not be necessary to work with a better approximation of the gradient.

    \begin{figure}
        \centering
        \includegraphics[width=0.45\textwidth]{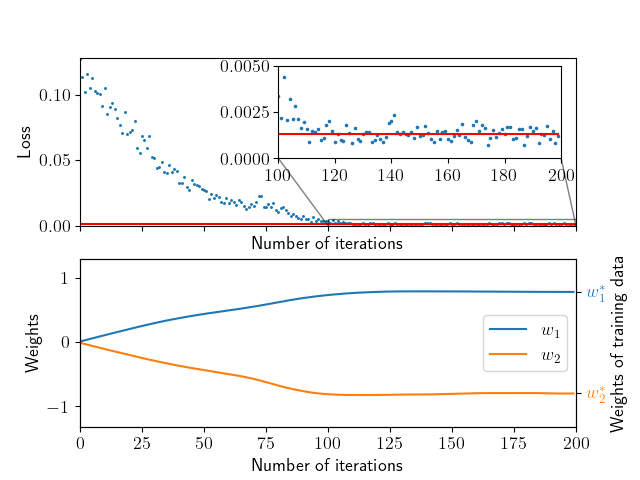}
        \caption{The training loss curve and the values of the weights during the learning process for the $3$-mode CVBM specified in Fig.~\ref{fig:circuits}. The blue dots indicate the losses at each iteration. The red line indicates a mean value of $L_{\MMD}(Q)$ averaged from $100$ values, where $Q$ is the target distribution.
        During the training, $1500$ shots were used to approximate the MMD loss and its gradient. The learning rate was $0.01$ and the Fock space cutoff was $10$.
        The training has converged at around iteration $100$.
        The $200$ iterations were executed in $870$ s on an Intel(R) Core(TM) i7-10610U CPU @ 1.80GHz architecture.}
        \label{fig:losses_weights_3_mode}
    \end{figure}
\begin{figure}[t]
        \centering
        \includegraphics[width=0.45\textwidth]{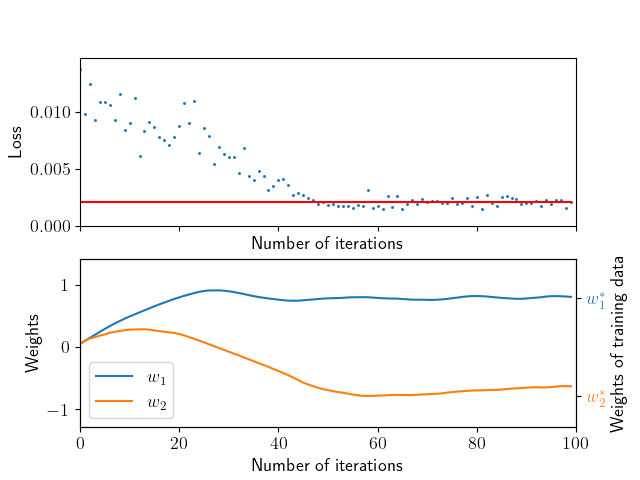}
        \caption{The training curve and the values of the weights during the learning process for the $4$-mode CVBM specified in Fig.~\ref{fig:circuits}.
        The blue dots on the training curve indicate each iteration's losses. The red line indicates a mean value of $L_{\MMD}(Q)$ averaged from $10$ values, where $Q$ is the target distribution.
        For the learning, $1000$ shots were used to approximate the MMD loss and its gradient. The learning rate was $0.05$ and the Fock space cutoff was $10$.
        The $100$ iterations were executed in $1455$ s on an Intel(R) Core(TM) i7-10610U CPU @ 1.80GHz architecture.            
        }
        \label{fig:losses_weights_4_mode}
    \end{figure}

    We also successfully trained similar $3$- and $4$-mode circuits visualized in Fig.~\ref{fig:circuits}. The target distribution was generated with weights
    $
        w_1^* = \pi / 4$ and $w_2^* = -\pi/4
    $
    in both cases.
    The training curve and the values of the weights during the training process are shown in Fig.~\ref{fig:losses_weights_3_mode} and in Fig.~\ref{fig:losses_weights_4_mode} for the $3$- and $4$-mode circuits, respectively.

\section{Conclusion and Outlook}
    In this paper, we have contributed to the study of continuous-variable Born machines using a new weak simulation algorithm for homodyne sampling.
    The execution time of this method scales naively as $\Tilde{\mathcal{O}}(e^{2 (d + c - 1) \log(d+c-1)})$ with $d$ being the number of modes and $c$ the Fock space cutoff. However, benchmarking this algorithm for increasing system size provides a heuristic scaling of execution time as $e^{1.2997 d}$ for fixed Fock space truncation. This numerical technique opens up an avenue for studying multimode simulation of homodyne measurements.
    In our investigations, we focused on the learning ability of CVBMs using stochastic MMD gradients, and we have successfully demonstrated training of $2$-, $3$- and $4$-mode CVBMs even in the experimentally more feasible stochastic gradient descent regime.

    Note that we only trained specific CVBM circuits where the parameter shift rule is applicable. An important progress in the field of photonic PQCs would be to find more general parameter shift rules or another experimentally feasible method for updating the circuit parameters.
    Note also that although the CVBM circuits showed convergence in the presented examples, such results, in general, may heavily depend on the circuit layout, the chosen target distribution, and the hyperparameters of the training.
    Hence, additional efforts are required to develop CVBMs that are more robust in this regard.
    Moreover, further research is needed to more precisely determine the sample-size region where the statistical fluctuations (or stochasticity) of the gradient values still allow for the convergence of CVBMs.

\newpage

\section*{Acknowledgment}
    We thank Zsófia Kallus and the Ericsson Research team for the inspiring discussions.
    This research was supported by the Ministry of Culture and Innovation and the National Research, Development and Innovation Office through the KDP-2021 funding scheme (Grant No. C1788111), Grants TKP2021-NVA-04 and FK135220.
    ZZ also acknowledges support from the QuantERA II project HQCC-101017733.

\onecolumn
\newpage
\twocolumn

\bibliographystyle{IEEEtran}
\bibliography{IEEEabrv,refs}

\end{document}